\newif\ifsubmit
\submitfalse

\newif\ifdraft
\draftfalse

\documentclass{sty/llncs/llncs}

\pdfoutput=1

\usepackage[T1]{fontenc}
\usepackage[utf8]{inputenc}
\usepackage{lmodern}
\usepackage[scaled=.8]{beramono}
\usepackage{latexsym}
\usepackage{textcomp}

\usepackage{amssymb}
\usepackage[final]{graphicx}
\usepackage{caption}
\usepackage[final]{listings}
\usepackage{sty/clange/lstsemantic}
\usepackage{booktabs}
\usepackage{mathtools}
\usepackage{tabularx}

\usepackage{graphicx}
\usepackage{rotating}
\usepackage{tabularx}
\usepackage{color}
\usepackage{multirow}

\definecolor{gray}{rgb}{0.4,0.4,0.4}
\definecolor{darkblue}{rgb}{0.0,0.0,0.6}
\definecolor{cyan}{rgb}{0.0,0.6,0.6}
\lstdefinelanguage{XML}
{
  morestring=[b]",
  morestring=[s]{>}{<},
  morecomment=[s]{<?}{?>},
  stringstyle=\color{black},
  identifierstyle=\color{darkblue},
  keywordstyle=\color{cyan},
  morekeywords={xmlns,version,type, title,dateofacceptance,publisher,resulttype,language,journal}
}

\usepackage[ngerman,british]{babel}

\usepackage[inline]{enumitem}
\usepackage{ctable}
\usepackage[babel]{csquotes}
\MakeAutoQuote{“}{”}

\usepackage{soul}
\setul{.25ex}{}

\usepackage[obeyDraft,textsize=tiny]{todonotes}

\usepackage{hyperref}

\begin{document}

\title{Mapping Large Scale Research Metadata to Linked Data: A Performance Comparison of HBase, CSV and XML%
}

\author{Sahar Vahdati\inst{1}
\and Farah Karim\inst{1}
\and Jyun-Yao Huang\inst{2}
\and Christoph Lange\inst{3}}
\institute{%
University of Bonn
\email{vahdati@uni-bonn.de}, \email{Karim@iai.uni-bonn.de}
\and National Chung Hsing University, Taiwan \email{allen501pc@gmail.com}
\and University of Bonn \& Fraunhofer IAIS, Germany
\email{math.semantic.web@gmail.com}
}

\maketitle

\begin{abstract}
OpenAIRE, the Open Access Infrastructure for Research in Europe, comprises a database of all EC FP7 and H2020 funded research projects, including metadata of their results (publications and datasets).
These data are stored in an HBase NoSQL database, post-processed, and exposed as HTML for human consumption, and as XML through a web service interface.
As an intermediate format to facilitate statistical computations, CSV is generated internally.
To interlink the OpenAIRE data with related data on the Web, we aim at exporting them as Linked Open Data (LOD).
The LOD export is required to integrate into the overall data processing workflow, where derived data are regenerated from the base data every day.
We thus faced the challenge of identifying the best-performing conversion approach.
We evaluated the performances of creating LOD by a MapReduce job on top of HBase, by mapping the intermediate CSV files, and by mapping the XML output.
\end{abstract}


\section{Introduction}
\label{sec:Intro}

\todo{CL: re-enable keywords if required}The European Commission emphasizes open access as a key tool to bring together people and ideas in a way that catalyses science and innovation. 
More than ever before, there is a recognized need for 
digital research infrastructures for all kinds of research outputs, across disciplines and countries.
OpenAIRE, the Open Access Infrastructure for Research in Europe (\url{http://www.openaire.eu}),
(1) manages scientific publications and associated scientific material via repository networks,
(2) aggregates Open Access publications and links them to research data and funding bodies, and
(3) supports the Open Access principles via national helpdesks and comprehensive guidelines.

Data related to those in the OpenAIRE information space exist in different places on the Web.
Combining them with OpenAIRE will enable new use cases.
For example, understanding changes of research communities or the emergence of scientific topics not only requires metadata about publications and projects, as provided by OpenAIRE, but also data about events such as conferences as well as a knowledge model of research topics and subjects (cf.~\cite{OsbMot:SemPub14}).

The availability of data that is free to use, reuse and redistribute (i.e.\ \emph{open data}) is the first prerequisite for analysing such information networks.
However, the diverse data formats and means to access or query data, the use of duplicate identifiers, and the heterogeneity of metadata schemas pose practical limitations on reuse.
Linked Data, based on the RDF graph data model, is now increasingly accepted as a lingua franca to overcome such barriers~\cite{scharffe2012}.

The University of Bonn is coordinating the effort of publishing the OpenAIRE data as Linked Open Data (LOD) and linking it to related datasets in the rapidly growing LOD Cloud\footnote{\url{http://lod-cloud.net}}.
This effort is further supported by the Athena Research and Innovation Center and CNR-ISTI.
Besides data about scientific events and subject classification schemes, relevant data sources include public sector information (e.g., to find research results based on the latest employment statistics, or to answer questions such as “how do the EU member states' expenses for health research compare to their health care spendings?”) and open educational resources (“how soon do emergent research topics gain wide coverage in higher education?”).

Concrete steps towards this vision are (1) mapping the OpenAIRE data model to suitable standard LOD vocabularies, (2) exporting the objects in the OpenAIRE information space as a LOD graph and (3) facilitating integration with related LOD graphs.
Expected benefits include
\begin{itemize}
\item enabling semantic search over the outputs of European research projects,
\item simplifying the way the OpenAIRE data can be enriched by third-party services, and consumed by interested data or service providers,
\item facilitated outreach to related open content and open data initiatives, and
\item enriching the OpenAIRE information space itself by exploiting how third parties will use its LOD graph.
\end{itemize}

The specifically tailored nature of the OpenAIRE infrastructure, its large amount of data (covering more than 11 million publications) and the frequent updates of the more than 5000 repositories from which the data is harvested pose high requirements on the technology chosen for mapping the OpenAIRE data to LOD.
We therefore compared in depth three alternative mapping methods, one for each source format in which the data are available: HBase, CSV and XML.

Section~\ref{sec:Inputdata} introduces the OpenAIRE data model and the three existing data sources.
Section~\ref{sec:vocab} presents our specification of the OpenAIRE data model as an RDF vocabulary.
Section~\ref{sec:Requirements} establishes requirements for the mapping.
Section~\ref{sec:SOTA} presents the state of the art for each of the three mapping approaches.
Section~\ref{sec:Implementation} explains our three implementations.
In section~\ref{sec:Evaluation} we evaluate them in comparison, with regard to different metrics induced by the requirements. 
Section~\ref{sec:RW} reviews work related to our overall approach (comparing mappings and producing research LOD).
Section~\ref{sec:Conclu} concludes and outlines future work.

\section{Input Data}
\label{sec:Inputdata}

\begin{figure}[h!b]
    \includegraphics[trim=0cm  28cm 22cm 41.3cm ,clip,width=\textwidth]{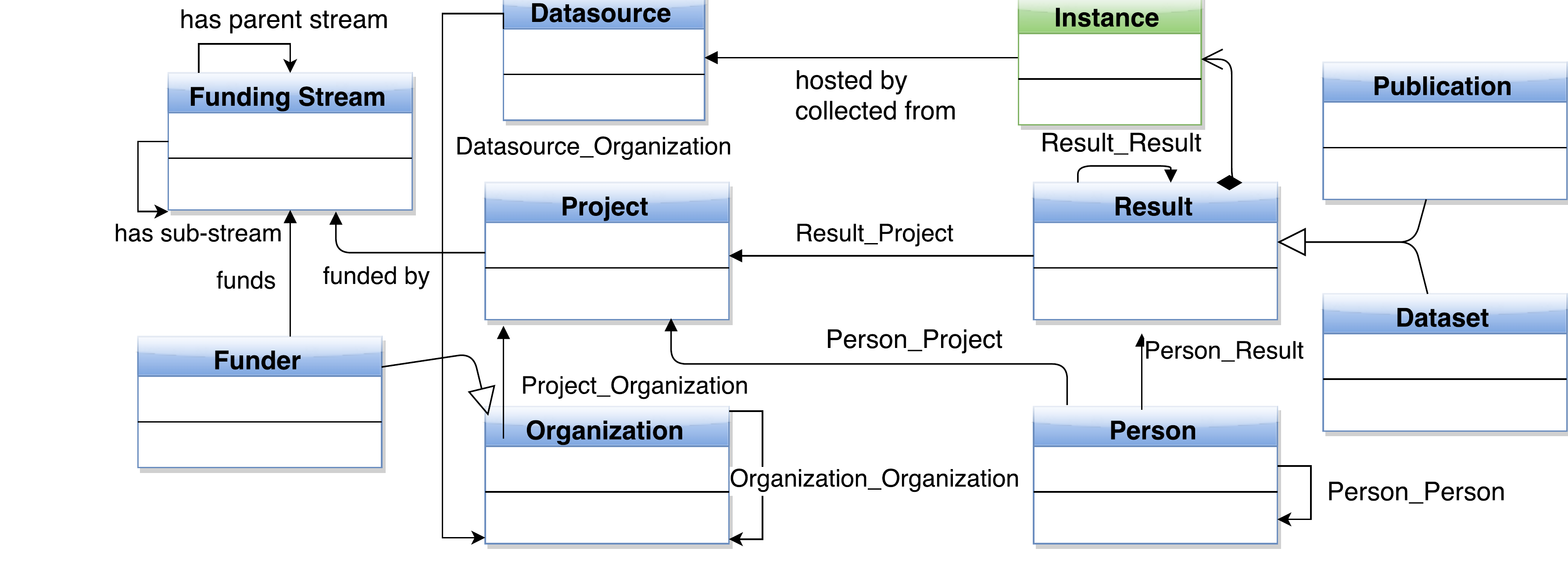}
    \caption{OpenAIRE Data Model: core entities and relationships}
    \label{fig:ER}
\end{figure}

The data model of OpenAIRE infrastructure is specified as an entity relationship model (ERM)~\cite{OpenAIREDataModel,manghi2012data} with the following entity categories:
\begin{itemize}
\item \textbf{Main entities} (cf.\ figure~\ref{fig:ER})\footnote{\url{https://issue.openaire.research-infrastructures.eu/projects/openaire2020-wiki/wiki/Core_Data_Model}}: Result (Publication or Dataset), Person, Organization, Projects, and DataSource (e.g.\ Repository, Dataset Archive or CRIS\footnote{Current research information system, a system to manage information about the research activity of an institution}).
Instances of these are continuously harvested from data providers.

\item \textbf{Structural entities} representing complex information about main entities: Instances (of a Result in different DataSources), WebResources, Titles, Dates, Identities, and Subjects.
 
\item \textbf{Static entities}, whose metadata do not change over time: Funding.
E.g., once a funding agency has opened a funding stream, it remains static.
 
\item \textbf{Linking entities} represent relationships between entities that carry further metadata; e.g., an entity of type Person\_Result whose property \textit{ranking} has the value 1 indicates the first author.
\end{itemize}

So far, the OpenAIRE data have been available in three formats: HBase, CSV and XML.

\subsection{HBase}
\label{sec:input-hbase}

Currently, the master source of all OpenAIRE data is kept in HBase, a column store based on HDFS (Hadoop Distributed File System).
HBase was introduced in 2012 when data integration efforts pushed the original PostgreSQL database to its limits: joins became inefficient and parallel processing, as required for deduplication, was not supported.
Each row of the HBase table has a unique row key and stores a main entity and a number of related linked entities.
The attribute values of the main entities are stored in the \textit{<family>:body} column, where the \textit{<family>} is named after the type of the main entity, e.g., \textit{result}, \textit{person}, \textit{project}, \textit{organization} or \textit{datasource}.
The attribute values of linked entities, indicating the relationship between main entities, are stored in dedicated column families \textit{<family>:<column>}, where \textit{<family>} is the class of the linked entity and \textit{<column>} is the row key of the target entity.
Both directions of a link are represented.
Cell values are serialized as byte arrays according to the Protocol Buffers~\cite{ProtocolBuffers} specification; for example:

\begin{lstlisting}[label=personSchema,
%caption=A Protocol Buffer Schema for Main Entity \textit{person},
morekeywords={message,optional,repeated}]
message Person {
	optional Metadata metadata = 2;
	message Metadata {
		optional StringField firstname = 1;
		repeated StringField secondnames = 2;
		optional Qualifier nationality = 9; ... }
	repeated Person coauthors = 4; }
\end{lstlisting}
  
The following table shows a publication and its authors.
For readability, we abbreviated row keys and spelled out key-value pairs rather than showing their binary serialization.
    
{\noindent
            \tiny
            \begin{tabular}{%
|>{\centering}m{1.3cm}
|>{\centering}m{2.0cm}
|>{\centering}m{2.3cm}
|>{\centering}m{2.0cm}|>{\centering}m{1.9cm}
|>{\centering\arraybackslash}m{2.0cm}
|} 
             \hline
          \multirow{2}{2cm}{RowKey} & result: & person: & \multicolumn{2}{c|}{\dots hasAuthor:} &   \dots isAuthorOf:\\ \cline{2-6}
            
            & body & body & 30\textbar\dots 001::9897\dots & 30\textbar\dots 001::ef29\dots  & 50\textbar\dots 001::39b9\dots \\ \hline
           
           50\textbar\dots 0 01::39 b9\dots & resulttype= ``publication''; title=``The Data Model of \dots''; dateofacceptance= ``2012-01-01''; language=``en''; publicationDate= ``2012''; publisher= ``Springer'';        & & ranking=1; & ranking=2;  & \\   \hline
          
           30\textbar \dots0 01::98 97\dots  &  & firstname=``Paolo''; lastname=``Manghi'';  & & & ranking=1; \\   \hline

           30\textbar \dots0 01::ef 29\dots & & firstname=``Nikos''; lastname=``Houssos''; & & & ranking=2; \\   \hline
            \end{tabular} 
}

\subsection{CSV}
\label{sec:input-csv}


CSV files aid the computation of statistics on the OpenAIRE information space.
HBase is a sparse key value-store designed for data with little or no internal relations.
Therefore, it is impossible to run complex queries directly on top of HBase, for example a query to find all results of a given project.
It is thus necessary to transform the data to a relational representation, which is comprehensible for statistics tools and enables effective querying.
Via an intermediate CSV representation, the data is imported into a relational database, which is queried for computing the statistics.

In this generation process, each main entity type (result, project, person, organization, datasource) is mapped to a CSV file of the same name, which is later imported into a relational database table.
Each single-valued attribute of an entity (id, title, publication year, etc.) becomes a field in the entity's table.
Multi-valued attributes, such as the publication languages of a result, are mapped to relation tables (e.g.\ \texttt{result\_languages}) that represent a one-to-many relation between entity and attributes.
Linked entities, e.g.\ the authors of a \textit{result}, are represented similarly.
As the data itself includes many special characters, for example commas in publication titles, the OpenAIRE CSV files use ! as a delimiter and wrap cell values into leading and trailing hashes:

\begin{lstlisting}
#dedup_wf_001::39b91277f9a2c25b1655436ab996a76b#!#The Data Model of the OpenAIRE
Scientific Communication e-Infrastructure#!#null#!#null#!#Springer#!#null#!#null
#!#null#!#null#!#2012#!#2012-01-01#!#Open Access#!#Open Access#!#Access#!#null#!#
0#!#null#!#nulloai:http://helios-eie.ekt.gr:!#publication#10442/13187oai:pumaoai.
isti.cnr.it:cnr.isti/cnr.isti/2012-A2-040#!#1#!
\end{lstlisting}

Finally, using CSV has the advantage that existing tools such as Sqoop can be used, thus reducing the need to develop and maintain customly implemented components on the OpenAIRE production system.

\subsection{XML}
\label{sec:input-xml}

OpenAIRE features a set of HTTP APIs\footnote{\url{http://api.openaire.eu/}} for exporting metadata as XML for easy reuse by web services.
These APIs use an XML Schema implementation of the OpenAIRE data model called OAF (OpenAIRE Format)\footnote{\url{https://www.openaire.eu/schema/0.2/doc/oaf-0.2.html}}, where each record represents one entity.
There is one API for searching, and one for bulk access.
For example, the listing below comes from \texttt{http://api.openaire.eu/search/publications\\?openairePublicationID=dedup\_wf\_001::39b91277f9a2c25b1655436ab996a76b} and shows the metadata of a publication that has been searched for.

\begin{lstlisting}[language=XML]
<oaf:result>
  <title schemename="dnet:dataCite_title" classname="main title"
   schemeid="dnet:dataCite_title" classid="main title">The Data Model of the
    OpenAIRE Scientific Communication e-Infrastructure</title>
  <dateofacceptance>2012-01-01</dateofacceptance>
  <publisher>Springer</publisher>
  <resulttype schemename="dnet:result_typologies" classname="publication"
   schemeid="dnet:result_typologies" classid="publication"/>
  <language schemename="dnet:languages" classname="English"
   schemeid="dnet:languages" classid="eng"/>
  <format>application/pdf</format>
  ...
</oaf:result>
\end{lstlisting}
The API for bulk access uses OAI-PMH (The \textbf{O}pen \textbf{A}rchives \textbf{I}nitiative \textbf{P}rotocol for \textbf{M}etadata \textbf{H}arvesting)\footnote{\url{http://www.openarchives.org/OAI/openarchivesprotocol.html}} to publish metadata and its corresponding endpoint is at \url{http://api.openaire.eu/oai_pmh}. 
The bulk access API lets developers fetch the whole XML files step by step.
For our experiments, we obtained the XML data directly from the OpenAIRE server, as an uncompressed Hadoop SequenceFile\footnote{\url{http://wiki.apache.org/hadoop/SequenceFile}} comprising 500 splits of $\sim$300 MB each.

\section{Implementing the OpenAIRE Data Model in RDF}
\label{sec:vocab}
As the schema of the OpenAIRE LOD we specified an RDF vocabulary by mapping the entities of the ER data model to RDF classes and its attributes and relationships to RDF properties.
We reused suitable existing RDF vocabularies identified by consulting the Linked Open Vocabularies search service\footnote{\url{http://lov.okfn.org}} and studying their specifications.
Reused vocabularies include
Dublin Core for general metadata, 
SKOS\footnote{\url{http://www.w3.org/2004/02/skos/}} for classification schemes and
CERIF\footnote{Common European Research Information Format; see \url{http://www.eurocris.org/cerif/main-features-cerif}} for research organizations and activities.
We linked new, OpenAIRE-specific terms to reused ones, e.g., by declaring \textit{Result} a superclass of \url{http://purl.org/ontology/bibo/Publication} and \url{http://www.w3.org/ns/dcat#Dataset}.

We keep the URIs of the LOD resources (i.e.\ entities) in the \url{http://lod.openaire.eu/data/} namespace.
We modelled them after the HBase row keys.
In OpenAIRE, these are fixed length identifiers of the form \{\textit{typePrefix}\}\textbar \{\textit{namespace\allowbreak Prefix}\}
::\textit{md5hash}. 
\textit{typePrefix} is a two digit code, 10, 20, 30, 40 or 50, corresponding to the main entity types datasource, organization, person, project and result.
The \textit{namespacePrefix} is a unique 12-character identifier of the data source of the entity.
For each row, \textit{md5hash} is computed from the entity attributes.
The resulting URIs look like \texttt{http://lod.openaire.eu/data/result/dedup\_wf\_001::39b9127}\\
\texttt{7f9a2c25b1655436ab996a76b}.

The following listing shows our running example in RDF/Turtle syntax.
\begin{lstlisting}[showstringspaces=false,language=N3,escapechar=!]
@prefix oad: <http://lod.openaire.eu/data/> .
@prefix oav: <http://lod.openaire.eu/vocab#> .
# further prefixes omitted; see !\url{http://prefix.cc}! for their standard bindings.
 
oad:result/...001::39b9... rdf:type oav:Result, bibo:Publication;
    dcterms:title "The Data Model of the OpenAIRE Scientific Communication
        e-Infrastructure"@en ;
    dcterms:dateAccepted "2012-01-01"^^xsd:date ;
    dcterms:language "en";
    oav:publicationYear 2012 ;
    dcterms:publisher "Springer";
    dcterms:creator oad:person/...001::9897..., oad:person/...001::ef29... .
oad:person/...001::9897... rdf:type foaf:Person;
    foaf:firstName "Paolo"; foaf:lastName "Manghi";
    oav:isAuthorOf oad:result/...001::39b9... .    
oad:person/...001::ef29... rdf:type foaf:Person;
    foaf:firstname "Nikos"; foaf:lastName "Houssos";
    oav:isAuthorOf oad:result/...001::39b9... .
\end{lstlisting}

\section{Requirements}
\label{sec:Requirements}

In cooperation with the other technical partners in the OpenAIRE2020 consortium, most of whom had been working on the infrastructure in previous projects for years, we established the following requirements for the LOD export:

\begin{enumerate}[label=R\arabic*]
\item The LOD output must follow the vocabulary specified in section~\ref{sec:vocab}.
\item The LOD must be generated from one of the three existing data sources, to avoid extra pre-processing costs.
\item The mapping to LOD should be maintainable w.r.t.\ planned extensions of the OpenAIRE data model (such as linking publications and data to software) and the evolution of linked data vocabularies.
\item\label{req:scalability} The mapping to LOD should be orchestrable together with the other existing OpenAIRE data provision workflows, always exposing a consistent view on the information space, regardless of the format.
\item To enable automatic and manual checks of the consistency and correctness of the LOD before its actual publication, it should be made available in reasonable time in a private space.
\end{enumerate}

To prepare an informed decision on the preferred input format to use for the LOD export, we realised one implementation for each of HBase, CSV and XML.

\section{Technical State of the Art}
\label{sec:SOTA}

For each possible approach, i.e.\ mapping HBase, CSV or XML to RDF, we briefly review the state of the art to give an overview of technology we could potentially reuse or build on, whereas section~\ref{sec:RW} reviews work related to our overall approach.
We assess reusability w.r.t.\ the OpenAIRE-specific requirements stated above.

\label{sec:hbase-rdf}%
\textbf{HBase}, being a sparse, distributed and multidimensional persistent sorted map, provides dynamic control over the data format and layout.
Several works have therefore explored the suitability of HBase as a triple store for semi-structured and sparse RDF data.
Sun et al. adopted the idea of the Hexastore indexing technique for storing RDF in HBase~\cite{sun2010scalable}.
Khadilkar et al. focused on a distributed RDF storage framework based on HBase and Jena to gain scalability
\cite{khadilkar2012jena}.
Others have provided MapReduce implementations to process SPARQL queries over RDF stored in HBase~\cite{papailiou2012h2rdf,haque2012distributed}.

We are only aware of one work on exposing data from column-oriented stores as RDF.
Kiran et al. provide a method for generating a SPARQL endpoint, i.e. a standardized RDF query interface, on top of HBase~\cite{hbaseRDF}.
They map tables to classes, rows to resources, and columns to properties.
Their approach do not scale well with increasing numbers of HBase entries, as the results show that the time taken to map HBase data to RDF is in hours for a few million rows~\cite{hbaseRDF}.

\label{sec:csv-rdf}%
\textbf{CSV} is widely used for publishing tabular data~\cite{lebo2010}.
The CSV on the Web W3C Working Group\footnote{\url{http://www.w3.org/2013/05/lcsv-charter.html}} provides technologies for data dependent applications on the Web working with CSV.
Several existing implementations, including that of Anything To Triples (any23)\footnote{\url{http://any23.apache.org}}, map CSV to a generic RDF representation.
Customizable mappings are more suitable for our purpose.
In Tarql (Transformation SPARQL)\footnote{\url{https://tarql.github.io}}, one can define such mappings in SPARQL; Tabels (Tabular Cells)\footnote{\url{http://idi.fundacionctic.org/tabels}} and Sparqlify\footnote{\url{https://github.com/AKSW/Sparqlify}~\cite{ermilov-ivan-2013-isem}} use domain-specific languages similar to SPARQL.
Tabels provides auxiliary machinery to filter and compare data values during the transformation process. 
Sparqlify is mainly designed to map relational databases to RDF but also features the sparqlify-csv module.

\textbf{XML} is used for various data and document exchange purposes.
Like for CSV$\to$RDF, there are generic and domain-specific XML$\to$RDF approaches.
Breitling implemented a direct, schema-independent transformation, which retains the XML structure
~\cite{breitling2009standard}.
Turning this generic RDF representation into a domain-specific one requires post-processing on the RDF side, e.g., transformations using SPARQL CONSTRUCT queries.
On the other hand, the current version 
of Breitling's approach is implemented in XSLT 1.0, which does not support streaming and is therefore not suitable for the very large inputs of the OpenAIRE setting.
Klein uses RDF Schema to map XML elements and attributes to RDF classes and properties~\cite{1046008}.
It does not automatically interpret the parent-child relation between two XML elements as a  property between two resources, but a lot of such relationships exist in the OpenAIRE XML.
XSPARQL can transform XML to RDF and back by combining the XQuery and SPARQL query languages to
~\cite{XSPARQL2012};
authoring mappings requires good knowledge of both.
By supporting XQuery's expressive mapping constructs, XSPARQL requires access to the whole XML input via its DOM (Document Object Model), which results in heavy memory consumption.
A subset of XQuery\footnote{cf. “Streaming in XQuery”, \url{http://www.saxonica.com/html/documentation/sourcedocs/streaming/streamed-query.html}} is suitable for streaming but neither supported by the XSPARQL implementation nor by the free version of the Saxon XQuery processor required to run XSPARQL.

\section{Implementation}
\label{sec:Implementation}

\label{sec:hbase-rdf-impl}
As the only existing \textbf{HBase$\to$RDF} implementation does not scale well (cf.\ section~\ref{sec:hbase-rdf}), we decided to follow the MapReduce paradigm for processing massive amounts of data in parallel over multiple nodes.
We implemented a single MapReduce job
.
Its mapper reads the attributes and values of the OpenAIRE entities from their protocol buffer serialization and thus obtains all information required for the mapping to RDF.
Hence no reducer is required.
The map-only approach performs well thanks to avoiding the computationally intensive shuffling.
RDF subjects are generated from row keys, predicates and objects from attribute names and cell values or, for linked entities, from column families/qualifiers.
 
\label{sec:csv-rdf-impl}
Mapping the OpenAIRE \textbf{CSV$\to$RDF} is straightforward: files correspond to classes, columns to properties, and each row is mapped to a resource.
We initially implemented mappings in Tarql, Sparqlify and Tabels (cf.\ section~\ref{sec:csv-rdf}) and ended up preferring Tarql because of its good performance\footnote{Tabels failed to handle large CSV files because it loads all the data from the CSV into main memory; Sparqlify works similar to Tarql but with almost doubled execution time (7,659 s) and more than doubled memory usage.} and the most flexible mapping language – standard SPARQL\footnote{\url{http://www.w3.org/TR/sparql11-query/}} with a few extensions. 
As we \emph{map} CSV$\to$RDF, as opposed to \emph{querying} CSV like RDF, we implemented \textit{CONSTRUCT} queries, which specify an RDF template in which, for each row of the CSV, variables are instantiated with the cell values of given columns.

\label{sec:xml-rdf-impl}
To enable easy maintenance of \textbf{XML$\to$RDF} mappings by domain experts, and efficient mapping of large XML inputs, we implemented our own approach\footnote{See source code and documentation at \url{https://github.com/allen501pc/XML2RDF}.}.
It employs a SAX parser and thus supports streaming.
Our mapping language is based on RDF triple templates and on the XPath\footnote{\url{http://www.w3.org/TR/xpath20/}} language for addressing content in 
XML
.
XPath expressions in the subjects or objects of RDF triple templates indicate where in the XML they obtain their values from.
To keep XPath expressions simple and intuitive, we allow them to be ambiguous, e.g., by saying that \textit{oaf:result/publisher/text()} (referring to the text content of the \textit{publisher} element of a result) maps to the \textit{dcterms:publisher} property of an \textit{oav:Result}, and that \textit{oaf:result/dateofacceptance/text()} maps to \textit{dcterms:date\allowbreak Accepted}.
In theory, any combination of \textit{publisher} and \textit{dateofacceptance} elements would match such a pattern; however in reality only those nodes that have the shortest distance in the XML document tree represent attributes of the \emph{same} OpenAIRE entity.
XML Filters~\cite{diao2002yfilter} efficiently restrict the XPath expressions to such combinations.




\section{Evaluation}
\label{sec:Evaluation}

\subsection{Comparison Metrics}

The \textbf{time} it takes to transform the complete OpenAIRE input data to RDF is the most important performance metric (requirement~\ref{req:scalability}).
The \textbf{main memory usage} of the transformation process is important because 
OpenAIRE2020 
envisages the development of further services sharing the same infrastructure, including deduplication, data mining to measure research impact, classification of publications by machine learning, etc.
One objective metric for \textbf{maintainability} is the size of the mapping's source code – after stripping comments and compression, which makes the comparison “independent of arbitrary factors like lengths of identifiers and amount of whitespace”~\cite{Wiedijk:tdbf:on}.\footnote{We used \texttt{tar cf - <input files> | xz -9}.  For HBase, we considered the part of the Java source code that is concerned with declaring the mapping, whereas our CSV and XML mappings are natively defined in high-level mapping languages.}
The “cognitive dimensions of notation” (CD) evaluation framework provides further criteria for systematically assessing the “usability of information artefacts”~\cite{CogDimNot}.
The following dimensions are straightforward to observe here: \emph{closeness} of the notation to the problem (here: mapping HBase/CSV/XML to RDF), \emph{terseness} (here measured by code size; see above), \emph{error-proneness}, 
\emph{progressive evaluation} (i.e. whether one can start with an incomplete mapping rule and evolve it to further completeness), and \emph{secondary notation and escape from formalism} (e.g. whether reading cues can be given by non-syntactic means such as indentation or comments).

\subsection{Evaluation Setup}

The \textbf{HBase$\to$RDF} evaluation ran on a Hadoop cluster of 12 worker nodes operated by CNR.\footnote{\url{https://issue.openaire.research-infrastructures.eu/projects/openaire/wiki/Hadoop_Clusters\#section-3}}
As our \textbf{CSV$\to$RDF} and \textbf{XML$\to$RDF} implementations required dependencies not yet installed there, we evaluated them locally: on a virtual machine on a server with an Intel Xeon E5-2690 CPU, having 3.7 GB memory and 250 GB disk space assigned and running Linux 3.11 and JDK 1.7.
As we did not have a cluster available, and as the tools employed did not natively support parallelization, we ran the mappings from CSV and XML sequentially.

\subsection{Measurements and Observations}

The following table lists our measurements; further observations follow below.

\noindent\begin{tabularx}{\textwidth}{Xlll}
     \FL
     Objective Comparison Metrics        & \textbf{HBase} & \textbf{CSV} & \textbf{XML} \ML
     Mapping Time(s)                     & 1,043          & 4,895        & 45,362        \NN
     Memory (MB)                         & 68,000         & 103          & 130          \NN
     Compressed Mapping Source Code (KB) & 4.9            & 2.86         & 1.67        \NN
     Number of Input rows/records        & 20,985,097     & 203,615,518  & 25,182,730     \NN
     Number of Generated RDF Triples     & 655,328,355    & 654,193,273  & 788,953,122     \LL
\end{tabularx}


For \textbf{HBase$\to$RDF}, the peak memory usage of the cluster was 68 GB, i.e. $\sim$5.5 GB per worker node.
No other MapReduce job was running on the cluster at the same time; however, the usage figure includes the memory used by the Hadoop framework, which schedules and monitors job execution.

The 20 \textbf{CSV} input files correspond to different entities but also to relationships.
This, plus the way multi-valued attributes are represented (cf.\ section~\ref{sec:input-csv}), causes the high number of input rows.
The size of all files is 33.8 GB.
The \textbf{XML$\to$RDF} memory consumption is low because of stream processing.
The time complexity of our mapping approach depends on the number of rules (here: 118) and the size of the input (here: 144 GB). 
With the complexity of the XML representation, this results in an execution time of more than 12 hours.
The size of the single RDF output file is $\sim$91 GB.
Regarding \emph{cognitive dimensions}, the different notations expose the following characteristics; for lack of space we focus on selected highlights.
\emph{Terseness}: the high-level CSV$\to$RDF and XML$\to$RDF languages fare better than the Java code required for HBase$\to$RDF.
Also, w.r.t. \emph{closeness}, they enable more intuitive descriptions of mappings.
As the CSV$\to$RDF mappings are based on SPARQL, which uses the same syntax for RDF triples than the Turtle RDF serialization, they look closest to RDF.
\emph{Error-proneness}: Syntactically correct HBase$\to$RDF Java code may still define a semantically wrong mapping.
In Tarql's CSV$\to$RDF mappings, many types of syntax and semantics errors can be detected easily. 
\emph{Progressive evaluation}: one can start with an incomplete Tarql mapping rule CSV$\to$RDF mapping rule and evolve it towards completeness.
\emph{Secondary notation}: Tarql and Java support flexible line breaks, indentation and comments, whereas our current XML$\to$RDF mapping implementation requires one (possibly long) line per mapping rule.
Overall, this strongly suggests that CSV$\to$RDF is the most maintainable approach.

\section{Related Work}
\label{sec:RW}

Comparisons of different approaches of mapping data to RDF have mainly been carried out for relational databases as a source~\cite{svihla2007, michel2013}.
Similarly to our evaluation criteria, the reference comparison framework of the W3C RDB2RDF Incubator Group 
covers 
mapping creation, representation and accessibility, 
and support for data integration
~\cite{sahoo2009}.
Hert et al.\ compared different RDB2RDF mapping languages 
w.r.t. syntactic features and semantic expressiveness~\cite{hert2011}.

For other linked datasets about research, we refer to the “publication” and “government” sectors of the LOD Cloud, which comprises, e.g., publication databases such as DBLP, as well as snapshots of funding databases such as CORDIS.
From this it can be seen that OpenAIRE is a more comprehensive data source than those published as LOD before.

\section{Conclusion and future work}
\label{sec:Conclu} 

We have mapped a recent snapshot of the OpenAIRE data to RDF.
A preliminary dump as well as the definitions of the mappings are available online at \url{http://tinyurl.com/OALOD}.
Mapping from HBase is fastest, whereas mapping from CSV promises to be most maintainable.
Its slower execution time is partly due to the less powerful hardware on which we ran it; comparing multiple CSV$\to$RDF processes running in parallel to the HBase$\to$RDF implementation on the CNR Hadoop cluster seems promising.
Based on these findings the OpenAIRE2020 LOD team will decide on the preferred approach for providing the OpenAIRE data as LOD; we will then make the data available for browsing from their OpenAIRE entity URIs, and for querying via a SPARQL endpoint.

Having implemented almost the whole OpenAIRE data model, future steps include interlinking the output with other existing datasets.
E.g., we so far output countries and languages as strings, whereas DBpedia and Lexvo.org are suitable linked open datasets for such terms.
Link discovery tools will further enable large-scale linking against existing “publication” and “government” datasets.


\paragraph{Acknowledgements.}
We would like to thank the partners in the OpenAIRE2020 project, in particular Claudio Atzori, Alessia Bardi, Glykeria Katsari and Paolo Manghi, for their help with accessing the OpenAIRE data.
This work has been partially funded by the European Commission under grant agreement no.\ 643410.


\end{document}